\documentclass[aps, pra, twocolumn, showpacs,footinbib,superscriptaddress]{revtex4-1}
\usepackage{graphicx, subfigure,braket}
\usepackage{amsmath,amssymb}
\usepackage{color}
\usepackage{natbib,hyperref}
\usepackage{placeins}
\usepackage{appendix}
\usepackage[T1]{fontenc}
\begin{document}

\title{Proposed real-time charge noise measurement via valley state reflectometry}

\author{David W. Kanaar}
\affiliation{Department of Physics, University of Maryland Baltimore County, Baltimore, MD 21250, USA}
\author{H.~Ekmel Ercan}
\affiliation{Department of Electrical and Computer Engineering,
University of California, Los Angeles, Los Angeles, CA 90095, USA}
\author{Mark F.~Gyure}
\affiliation{Department of Electrical and Computer Engineering,
University of California, Los Angeles, Los Angeles, CA 90095, USA}
\affiliation{Center for Quantum Science and Engineering, University of California, Los Angeles, Los Angeles, CA 90095, USA}
\author{J.~P.~Kestner}
\affiliation{Department of Physics, University of Maryland Baltimore County, Baltimore, MD 21250, USA}

\begin{abstract}
We theoretically propose a method to perform \textit{in situ} measurements of charge noise during logical operations in silicon quantum dot spin qubits. Our method does not require ancillary spectator qubits but makes use of the valley degree of freedom in silicon. Sharp interface steps or alloy disorder in the well provide a valley transition dipole element that couples to the field of an on-chip microwave resonator, allowing rapid reflectometry of valley splitting fluctuations caused by charge noise. We derive analytic expressions for the signal-to-noise ratio that can be expected and use tight binding simulations to extract the key parameters (valley splitting and valley dipole elements) under realistic disorder. We find that unity signal-to-noise ratio can often be obtained with measurement times below $1$ms, faster than typical decoherence times, opening the potential for closed-loop control, real-time recalibration, and feedforward circuits.
\end{abstract}

\maketitle
\section{Introduction}
Semiconductor spin qubits are a promising platform for building scalable quantum computing devices because of their compatibility with industrial CMOS technology~\cite{maurand_cmos_2016,gonzalez-zalba_scaling_2021}. Additionally, single- and two-qubit device fidelities high enough for error-correcting codes have been reached in semiconductor devices~\cite{yoneda_quantum-dot_2018,yang_silicon_2019,huang_fidelity_2019,petit_universal_2020,xue_quantum_2022,mills_two-qubit_2022,noiri_fast_2022,huang_high-fidelity_2023}, although full implementation of standard error-correction codes would require orders of magnitude more qubits than are available in current devices. 

Charge noise is an important factor that currently hampers performance at the level of the physical spins by causing fluctuations in the Larmor frequency through the spin-orbit interaction~\cite{ruskov_electron_2018}, as well as in the exchange coupling between spins~\cite{culcer_dephasing_2009,Keith2022}. Reducing the effect of charge noise at the physical spin level would also reduce the number of spin qubits required per encoded logical qubit.
The power spectral density of charge noise in these types of devices has been characterized via transport through a sensor dot or proximal quantum point contact~\cite{freeman_comparison_2016,yoneda_quantum-dot_2018,connors_low-frequency_2019,struck_low-frequency_2020,connors_charge-noise_2022,esposti_low_2023} and consistently shows a $1/f$ frequency dependence. It would be useful to track the low-frequency behavior of this noise \emph{during} the operation of a quantum logic circuit to achieve higher gate fidelities. In this paper we propose a method to do so.

Recent work has considered repeated quantum measurements (such as gate set tomography~\cite{Nielsen2021}) to experimentally optimize high-fidelity gates~\cite{Cerfontaine2020,Feng2018}. Such protocols generally use projective measurements to do an advance tune-up of control fields before starting a quantum protocol. The tomography stage is a slow process, on the order of hours, so it is only effective against equally slow drifts. It is a good way to pre-calibrate, but it is not a real-time recalibration scheme. 

Other calibration procedures have been explored that shortcut the full complexity of gate tomography. For instance, when one has a good model to begin with, it is more efficient to use real-time Hamiltonian estimation~\cite{Shulman2014,huang_fidelity_2019,Nakajima2020,Kim2022,vepsalainen_improving_2022}, which has successfully been used in spin qubit systems to calibrate away nuclear spin Overhauser noise. However, even these approaches still rely on repeated projective quantum measurement. Since this is currently the most expensive resource in spin qubit devices, it is attractive to consider even more efficient protocols using classical measurements (i.e., measurements of classical quantities). 

Classical measurement has previously been used to calibrate certain control fields, such as to automate the tune-up of the top gate voltages defining a quantum dot structure to the desired charge occupancy regime~\cite{Teske2019}. Fast classical feedback control has also been used to stabilize a charge sensor dot at an optimally sensitive point using a proportional-integral-derivative (PID) controller on a field programmable gate array (FPGA)~\cite{Nakajima2021}. There the sensor data was used to stabilize the \emph{sensor} dot itself, which is not the same as stabilizing the \emph{qubit} dot. For silicon-based devices with overlapping top gates, the degree of noise correlation between the sensor dot and qubit dot is still unclear.  Furthermore, when stabilizing a sensor dot in real-time, one can directly monitor its local chemical potential via the conductance through the dot~\cite{Nakajima2021}. For a qubit dot, the electron must remain localized to the dot for the duration of the quantum operations, and no conductance is allowed except during initialization or readout.

The key to using classical measurements during quantum operations is to measure an auxiliary degree of freedom that is independent of the qubit degree of freedom but couples to the same noise source. This has been considered in the context of introducing nearby ``spectator qubits'' which are periodically measured mid-circuit~\cite{Singh2023,majumder2020real,Gupta2020}. The auxiliary degree of freedom should be sensitive to the noise source but have a weak relaxation rate so that the noise information is not lost before it can be measured. 

In this paper we propose a method to carry out \textit{in situ} measurements of charge noise-induced electric field fluctuations in real time, i.e., the information is extracted without collapsing the qubit states and can be used for recalibration or feedforward during an ongoing run of a logic circuit. Our proposal is similar in spirit to that of mid-circuit measurement of spectator qubits except that in our case the spectator lives on the same physical electron as the qubit, which is ideal from a resource standpoint as well as a qubit-sensor correlation standpoint.

For an electron in a silicon quantum dot, the spin provides a good qubit degree of freedom while the valley has the desired properties of the auxiliary sensor degree of freedom. The valley states of bulk silicon are six-fold degenerate, but for electrons confined in a strained silicon quantum well or dot this degeneracy is reduced to twofold. The ground and excited valley states are a superposition of these two lowest energy valley states and the energy difference between them is called the valley splitting. Valley lifetimes as long as 12 ms have been observed in experiment~\cite{penthorn_direct_2020}. This is several orders of magnitude longer than orbital excitation lifetimes~\cite{tahan_relaxation_2014}, and is the reason we focus on the valley degree of freedom instead of orbital excitations. A resonator can weakly couple to the intervalley transition via disorder. Attractively, both qubit and sensor are colocated in a single electron.

Interface roughness and alloy disorder induce an electric transition dipole moment between the valley states of a single electron in a quantum dot without inducing a dipole between the spin states. The dipole moment arises from slightly hybridizing the orbital states with the valley states. However, the degree of hybridization depends sensitively on the details of the interface roughness and alloy disorder. As is common in the literature, we will refer to the lowest pair of these hybridized eigenstates simply as valley states. As a result of the induced dipole moment, a resonator will couple preferentially to the valley states while only weakly coupling to spin. Note that this differs from the method of previous dispersive measurements of valley splitting~\cite{burkard_dispersive_2016,mi_high-resolution_2017} in that the coupling to the resonator is due to disorder within the dot rather than tunnel coupling to a second quantum dot.

Through effective mass and atomic tight-binding theory~\cite{ruskov_electron_2018}, it has been shown that the valley splitting depends on the electric field at the dot. Charge noise induces fluctuations in this electric field. Thus, any charge noise induced fluctuation in the valley splitting can be detected dispersively via homodyne microwave reflectometry measurements near the resonance point of the coupled valley-resonator system. Electric field fluctuations can be inferred without collapsing the spin state, opening an avenue for closed-loop control of the gate voltages to compensate for the part of the charge noise that is slow compared to the measurement time.

This paper is divided as follows: In Sec.~\ref{sec:mastereq} we describe the system Hamiltonian and master equation. Then in Sec.~\ref{sec:inputoutputtheory}, we use input-output theory to find analytical expressions for the resonance frequency and the sensitivity of the out-of-phase voltage to fluctuations in valley splitting, confirmed by numerical master equation simulations in Appendix~\ref{sec:NumericalMaster}. In Sec.~\ref{sec:SNR}, we also derive the signal-to-noise ratio (SNR) and measurement time achievable from valley splitting fluctuations caused by charge noise. In Sec.~\ref{sec:tightbinding} we use tight binding simulations to find realistic dipole moments and valley splittings. Combining these numbers with experimental device parameters allows us to find measurement times where unity SNR is achieved in the parameter regime where input-output theory holds. Finally, we conclude in Sec.~\ref{sec:conclusions}.

\section{Coupled resonator-valley system}\label{sec:mastereq}
\begin{figure} 
    \centering 
    \includegraphics[width=0.9\linewidth]{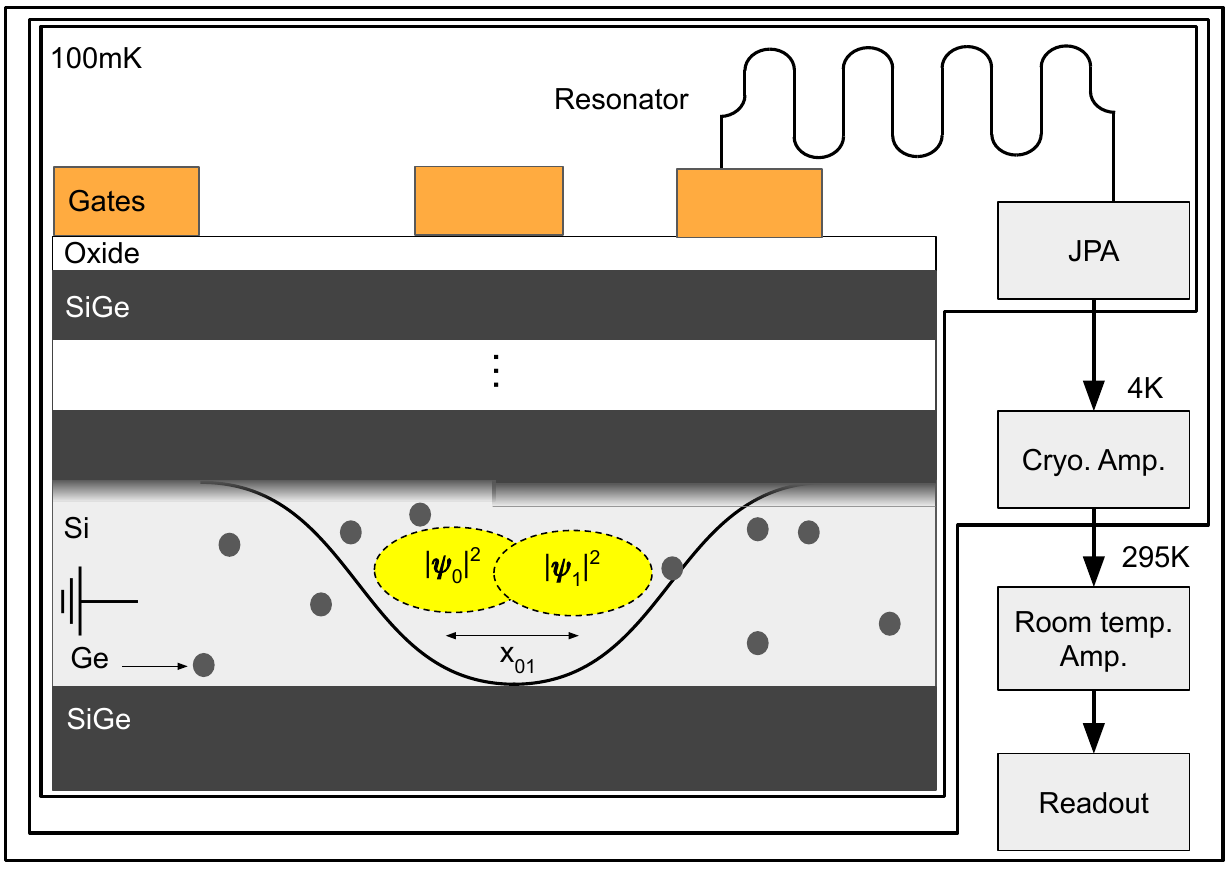}
    \caption{Schematic of the device and amplification chain used measurement. The silicon quantum well is randomly doped with 5\% Ge atoms and transitions into SiGe. The charge densities of the valley states, $|\psi|^2$, with an intervalley dipole moment  $x_{01}$, are coupled to the resonator through the metal gate.}  \label{fig:system}
\end{figure}
The Hamiltonian of a two-level system, the low-lying valley states in this case depicted in Fig.~\ref{fig:system}, coupled to a resonator in the rotating frame, $U_R=e^{i \omega_p t(a^{\dagger}a+\frac{1}{2}\sigma_Z)}$, with a probe frequency, $\omega_p$, after the rotating wave approximation is
\begin{equation}
    H_{sys}=(\omega_c-\omega_p)a^{\dagger}a + \frac{(\omega_{vs}-\omega_p)}{2}\sigma_z + g(a\sigma_{+}+a^{\dagger}\sigma_{-}),
\end{equation}
where $\omega_c$ is the resonator frequency, $\omega_{vs}$ is the valley splitting, $g$ is the dipole-induced coupling strength between the resonator and the valley state, $a$ is the lowering operator of the resonator, $\sigma_{\pm}=(\sigma_x\pm i\sigma_y)/2$, and $\sigma_{x,y,z}$ are Pauli operators. The coupling between the resonator and the valley states comes from an intervalley dipole moment, $x_{01} e$. The strength of the coupling is $g=E_{rms} x_{01} e/\hbar$~\cite{blais_cavity_2004}, where $E_{rms}$ is the root-mean-square (rms) of the zero point fluctuating field of the resonator in the dot. Defects in the boundary layers of the quantum well or random Ge atoms in the well can cause the valley states to have a dipole that would not exist if the well and boundary layer were perfect. Dipole moments corresponding to $x_{01}=50$nm have been measured~\cite{freeman_comparison_2016}. We will calculate an appropriate value for our proposal via tight-binding simulations in Sec.~\ref{sec:TBsims}.

Leakage out of the system is described using the Lindblad operator $\frac{\kappa}{2} L[a]$ where $\kappa$ is the cavity decay rate and 
\begin{equation}
    L[A]=(2A \rho A^{\dagger}-A^{\dagger} A \rho-\rho A^{\dagger} A).
\end{equation} 
Similarly, the excited valley decay rate is described by $\frac{\gamma}{2} L[\sigma_{-}]$ where $\gamma$ is the excited valley state decay rate. The resonator is coupled to an external probe field, resulting in an additional term in the Hamiltonian 
\begin{equation}\label{eq:ham}
    H = H_{sys}+i \sqrt{\kappa} b(a-a^{\dagger}),
\end{equation}
where $b$ is the probe field bath operator~\cite{mutter_fingerprints_2022}. The probe field is taken to be a single-frequency standing wave, which means in the rotating frame $b$ is a constant. The resulting evolution of the density matrix, $\rho$, as determined by the master equation is
\begin{equation}\label{eq:mastereqn1}
    \frac{d \rho}{dt} = - i [H,\rho] + \frac{\kappa}{2} L[a] +\frac{\gamma}{2} L[\sigma_{-}].
\end{equation}

For the highest sensitivity of the signal to fluctuations in valley splitting we examine the resonance point, $\omega_p=\omega_{res}$, where the output signal is at a maximum and in phase with the input signal. In other words, at resonance the out-of-phase voltage, $V_Q$, of the coupled transmission line of resistance $R$ is~\cite{blais_cavity_2004}
\begin{equation}\label{eq:VQ}
    V_Q = \sqrt{\hbar \omega_c \kappa R/2} \left\langle a-a^{\dagger}\right \rangle=0.
\end{equation} 
Any changes to the valley splitting result in a non-zero voltage through changes in the value of $\langle a \rangle$. Solving the master equation analytically to find $\langle a(t) \rangle$ is not possible. Therefore, to simplify the problem we assume the ratio of the frequency difference, $\Delta=\omega_{vs}-\omega_c$, to the coupling $g/\Delta\ll 1$ is in the dispersive regime. Then input-output theory~\cite{collett_squeezing_1984,gardiner_input_1985} can be used to simplify the problem. The values of $g$ and $\Delta$ calculated from detailed simulations in Sec.~\ref{sec:TBsims} respect this approximation and numerical master equation simulations in Appendix \ref{sec:NumericalMaster} confirm the validity of input-output theory.

\section{signal sensitivity to electric field fluctuations}\label{sec:inputoutputtheory}
We first show the effect of virtual valley excitations on the resonator signal near resonance. Then we argue that the effect of virtual orbital excitations merely shifts the resonance point and only has a marginal (and beneficial) effect on the sensitivity to the electric field. However, there is some loss of signal due to real orbital excitations followed by phonon decay.

\subsection{Effect of virtual valley excitations}
Here we find analytical expressions for the resonance frequency, $\omega_{res}$, as well as the sensitivity of the out-of-phase voltage to valley splitting at the resonance frequency,
\begin{equation}\label{eq:sensitDef}
\mathcal{S}\equiv\frac{dV_Q}{d\omega_{vs}}\bigg\rvert_{\omega_p=\omega_{res}}.
\end{equation}
The time evolution of the expectation values of operators in the master equation formalism is
\begin{equation}
   \frac{ d \left \langle A\right \rangle}{dt}= i \left \langle[H, A]\right \rangle+\left \langle\frac{\kappa}{2}D[A,a^{\dagger}]\right \rangle+\left \langle\frac{\gamma}{2}D[A,\sigma_{+}]\right \rangle
\end{equation}
where $D[A,b]=2 b A b^\dagger-\{b b^\dagger,A \}$.
The differential equations for $\langle a \rangle$ and the operators coupled to it are
\begin{align}
    \frac{ d\left \langle a\right \rangle}{dt} &=(-i(\omega_c-\omega_p)-\frac{\kappa}{2}) \langle a\rangle - i g \langle \sigma_- \rangle + \sqrt{\kappa} b \label{eq:dadt} \\
    \frac{d \left \langle a^{\dagger}\right \rangle}{dt}&=(i(\omega_c-\omega_p)-\frac{\kappa}{2}) \langle a^{\dagger}\rangle + i g \langle \sigma_+ \rangle + \sqrt{\kappa} b \\
    \frac{d \left \langle \sigma_z\right \rangle}{dt}&=2 i g ( \langle a\rangle \langle \sigma_{+} \rangle+  \langle a^{\dagger}\rangle \langle \sigma_{-} \rangle)-\gamma(1+\langle \sigma_{z}\rangle)\\
    \frac{d \left \langle \sigma_{+}\right \rangle}{dt}&=(i(\omega_{vs}-\omega_p)-\frac{\gamma}{2}) \langle \sigma_{+}\rangle+ig\langle a^{\dagger}\rangle\langle \sigma_{z}\rangle \\
    \frac{ d\left \langle \sigma_{-}\right \rangle}{dt}&=(-i(\omega_{vs}-\omega_p)-\frac{\gamma}{2}) \langle \sigma_{-}\rangle-ig\langle a\rangle\langle \sigma_{z}\rangle, \label{eq:dsigma-dt}
\end{align}
where we have assumed that in this dispersive limit $\langle O_A O_B\rangle = \langle O_A \rangle \langle O_B\rangle +\mathcal{O}(g)$. The results using this approximation below are confirmed by comparisons to numerical solutions of the master equation in Appendix~\ref{sec:NumericalMaster}.

We found analytical steady-state solutions to the coupled equations above, but the resulting expressions are too complex to gain insight from. The expressions simplify if we assume the steady-state $\langle \sigma_{z} \rangle_{ss}\approx-1$, but that is too crude of an approximation as it misses a $b$-dependence (i.e., an input power dependence) of the resonance frequency $\omega_{res}$ that we see in numerical master equation simulations. A more accurate simplification is to Taylor series expand the full expression for $\langle \sigma_z\rangle_{ss}$ in $b/\sqrt{\kappa}$ around 0 while also taking $\gamma \rightarrow 0$, which is justified in practice since the valley relaxation rate is much slower than any other frequency scale in the problem. The resulting expression is
\begin{equation} \label{eq:steadysigmaz}
    \begin{split}
        \langle \sigma_z\rangle_{ss} &\approx \frac{8 g^2 b^2\kappa}{4(g^2-(\omega_c-\omega_p)(\omega_{vs}-\omega_p))+(\omega_{vs}-\omega_p)^2\kappa^2}-1 \\ %+\mathcal{O}(b^4)
    &\approx  \frac{8 b^2 g^2 }{\kappa \Delta^2 }-1,
    \end{split}
\end{equation}
where in the final line we take $\omega_p$ to be close to the dispersive limit, $(\omega_c-\omega_p)\approx g^2/\Delta$ and $(\omega_{vs}-\omega_p)\approx \Delta$. 

We find the resonance frequency, $\omega_{res}$, in terms of $\langle \sigma_z\rangle_{ss}$ by solving $\text{Im}(\langle a \rangle_{ss})=0$ at $\gamma=0$ for $\omega_p$. There is a solution at $\omega_p=\omega_{vs}$, but the solutions we are interested in are determined by
\begin{equation} \label{eq:omegap}
    \omega_p-\omega_c+\langle\sigma_z\rangle \frac{g^2}{\omega_{vs}-\omega_p}=0.
\end{equation}
The lowest frequency solution to Eq.~\eqref{eq:omegap} is
\begin{equation}\label{eq:omegadispexpan}
    \begin{split}
        \omega_p&=\frac{1}{2} \left(\omega_c+\omega_{vs}-\Delta\sqrt{1-\frac{4g^2}{\Delta^2}\langle\sigma_z\rangle}\right)\\
        &\approx \omega_c+\frac{g^2 \langle\sigma_z\rangle}{\Delta}+\frac{g^4\langle\sigma_z\rangle^2}{\Delta^3}+ \Delta\mathcal{O}\left(\frac{g^{12} \langle\sigma_z\rangle^6}{\Delta^{12}}\right),
    \end{split}
\end{equation}
where we have expanded in the smallness parameter $g/\Delta$.
Substituting the solution for $\langle\sigma_z\rangle$ from Eq.~\eqref{eq:steadysigmaz} into Eq.~\eqref{eq:omegadispexpan} and keeping up to second order in $b/\sqrt{\kappa}$ and fourth order in $g/\Delta$ the final resonance frequency, $\omega_{res}$, is
\begin{equation} \label{eq:omegares}
    \omega_{res}=\omega_c-\frac{g^2}{\Delta}+\frac{g^4}{\Delta^3}\left(1+ \frac{8b^2}{\kappa}\right).
\end{equation}

Using this resonance frequency we find an analytical expression for the sensitivity of the out-of-phase voltage, $\mathcal{S}$ as defined in Eq.~\eqref{eq:sensitDef}, to changes in the valley splitting. This is done by calculating $\langle a \rangle_{ss}$ by combining Eqs.~\eqref{eq:dadt}, \eqref{eq:dsigma-dt}, and \eqref{eq:steadysigmaz}, which in turn yields $V_Q(\omega_{vs})$ from Eq.~\eqref{eq:VQ}. Then we take the derivative with respect to valley splitting. This produces an algebraically complicated expression, but in the limit where $\Delta$ is much larger than $g$, $\kappa$, $\gamma$ and $b^2$, as is assumed here, $\mathcal{S}$ is well approximated by its Taylor series in $1/\Delta$ around 0 as
\begin{equation}
\label{eq:sense}
    \mathcal{S}\approx4\sqrt{2 R \hbar \omega_c}\frac{b}{\kappa} \left(-\frac{g^2}{\Delta^2}+\frac{g^4}{\Delta^4}\left( 2+\frac{24b^2}{\kappa}\right) \right).%+ \mathcal{O}(\frac{1}{\Delta})^5 %\equiv\frac{d V_Q}{d\omega_{vs}}\Bigr\rvert_{\omega_p=\omega_{res}}
\end{equation}
For resonant probing, $\omega_p = \omega_{res}$, the number of photons in the resonator, $n_{res}$, is the square of $\text{Re}(\langle a\rangle_{ss})$ since $\text{Im}(\langle a\rangle_{ss})=0$. In the same limit of large $\Delta$, $n_{res}$ is also well approximated by
\begin{equation}\label{eq:napprox}
    n_{res} \approx \frac{4 b^2}{\kappa}.
\end{equation}
\subsection{Effects of orbital excitations}\label{sec:orbitalseffects}
In addition to the dispersive shift of Eq.~\eqref{eq:omegares} due to the intervalley dipole, there is an additional dispersive shift due to the resonator-orbital coupling, 
\begin{equation}
    \delta_{res}=-\frac{g_{o}^2}{\Delta_o}+\frac{g_o^4}{\Delta_o^3}\left(1+ \frac{8b^2}{\kappa}\right),
\end{equation}
where $g_o$ is the orbital coupling rate and $\Delta_o=\omega_{o}-\omega_c$ is the difference between the orbital and resonator frequencies. However, unlike the shift due to the resonator-valley coupling, this one is not sensitive to the electric field since the out-of-plane field does not affect the low-lying transitions and the in-plane field only shifts the dot position (approximating the lateral confinement potential as parabolic) and does not change the confinement energy directly. Although changes in interface and well disorder over this position shift do allow the orbital energy to change slightly, which would help enhance the sensitivity, this is expected to be a small effect compared to the electric field dependence of the valley splitting and consequently we neglect it.

The remaining effect of the existence of orbital transitions is to reduce the signal through real off-resonant photon absorption followed by rapid decay to phonons. The effect on the sensitivity of the out-of-phase voltage to valley splitting fluctuations is quantified in the Appendix~\ref{sec:NumericalMaster} where we did numerical master equation simulations taking into account the orbital levels by adding to the Hamiltonian of Eq.~\eqref{eq:ham} orbital terms 
\begin{equation}\label{eq:hamorb}
    H_{o}=\frac{\omega_{o}-\omega_p}{2} \tau_z+g_o(a \tau_{+}+a^\dagger \tau_-)
\end{equation}
and adding a decay term of $\frac{\gamma_o}{2}L[\tau_-]$ to the master equation of Eq.~\eqref{eq:mastereqn1}, where $\tau_{\pm}=(\tau_x\pm i \tau_y)/2$ and $\tau_{x,y,z}$ are the Pauli operators acting on the orbital subspace and the orbital decay rate, $\gamma_{o}$, in Si/SiGe dots is around 1-10GHz~\cite{tahan_relaxation_2014,raith_theory_2011,langrock_blueprint_2023}.
Depending on our choice of numbers, the sensitivity was decreased by between 4-30\%. Since we are simply interested in estimating the order of magnitude of the measurement time required for unity SNR, we neglect this effect in the remainder of our analysis for simplicity. Furthermore, we find the excited orbital occupation is only about 0.1, so the dissipated power is $0.1 \omega_o \gamma_o < 1$ pW as shown in Appendix~\ref{sec:NumericalMaster}. This is negligible compared to a cooling power of 2 mW at 100 mK~\cite{zu_development_2022}. 

\section{Signal to noise ratio}\label{sec:SNR}
For the $V_Q$ signal to be detectable the SNR must be at least unity. We consider the amplification chain shown on the right side of Fig.~\ref{fig:system}, which includes a Josephson parametric amplifier (JPA). The gain available from a JPA depends on the bandwidth required, but in our application a bandwidth of only 1 MHz is sufficient, so we assume a gain of $30$ dB~\cite{qiu_broadband_2023} with quantum-limited noise of half a photon. The amplifier chain also includes a 4 K cryogenic amplifier with a gain of 30 dB and a room-temperature amplifier with a gain of 40 dB, for a total gain of $G=100$ dB. 

An additional source of noise is the backaction of the dispersively coupled valley on the resonator leading to a broadening of the resonator linewidth. This effect can be modeled as an increase in $\gamma$, leading to an effective decay rate $\gamma_{eff}=\gamma+\kappa \sqrt{2 n}$~\cite{ibberson_large_2021} which does not affect the order of magnitude measurement time calculations as shown in Appendix~\ref{sec:appenFourthorder}. 

Finally, shot noise in the number of photons in the resonator results in an uncertainty in the output power $P_{out}=\hbar \omega_c \kappa \langle n\rangle/2$. The correlation function of the number of photons in the cavity which describes this noise is~\cite{blais_cavity_2004} 
\begin{equation}
    \langle n(0), n(t)\rangle= n e^{-\kappa |t|/2}.
\end{equation}
The Fourier transform of this is the power spectral density of the photon shot noise, 
\begin{equation}
    S_{nn}= \frac{4 n \kappa}{\kappa^2+f^2}.
\end{equation}
For a measurement time, $t_m$, this results in a variance in the number of photons,
\begin{equation}
    \sigma_n^2=\int_{0}^{1/t_m} S_{nn}=4 n \, \text{arccot}(\kappa t_m),
\end{equation}
and a rms noise power,
\begin{equation} \label{eq:noisepower}
    N = G \hbar \omega_c \kappa \sqrt{ n \, \text{arccot}(\kappa t_m)}.
\end{equation}
For the parameters obtained in Sec.~\ref{sec:tightbinding} $n_{\text{crit}}=\Delta^2/4g^2$ is a large number ($10^3$-$10^5$) which means the noise is dominated by photon shot noise and well approximated by Eq.~\eqref{eq:noisepower}. 
The signal power of the change in $V_Q$ as a result of charge noise is
\begin{equation}
    S=G (\mathcal{S} \vec{\nabla}_F \omega_{vs} \cdot \vec{\delta F})^2/R
\end{equation}
where the $\vec{\delta F}$ is the electric field fluctuation as a result of charge noise and $\vec{\nabla}_F \omega_{vs}$ is the gradient of the valley splitting with respect to the electric field. 
The sensitivity, $\mathcal{S}$, depends on the number of photons in the resonator. Using the approximation for the number of photons when driving on resonance from Eq.~\eqref{eq:napprox}, $n_{\text{res}}=4b^2/\kappa$, we choose the probe driving strength to be $b=\sqrt{ \kappa}\Delta/30g$ such that $n= \frac{2}{15}n_{\text{crit}}$ where the SNR is maximized. Equation \eqref{eq:sense} then becomes 
\begin{equation}
   \mathcal{S}\approx-\sqrt{\frac{ \hbar \omega_c R}{15 \kappa}} \left(\frac{8}{5}\frac{g}{\Delta}\right)
\end{equation}
after dropping the $(g/\Delta)^3$ term.

The size of the electric field fluctuation is determined by the magnitude of the charge noise. Experimentally, charge noise is often characterized in terms of the resulting chemical potential fluctuations, $\delta\mu$, of the dot. This does not encapsulate all information about the actual noise, such as the direction of the field fluctuation, but it does provide an overall magnitude. This can be converted to a measure of the gate-referred voltage fluctuation~\cite{Reed2016} through the dot lever arm, $\alpha$, and then assuming the voltage varies approximately as the inverse of the distance from the gate, $l$, the electric field fluctuation at the dot location is given by $\delta F=\delta \mu /\alpha l.$
For charge noise power spectral density of the form $A_0/f$ up to a high-frequency cutoff, $f_{\text{cutoff}}$, integrating over the frequency range of $1/t_m$ up to the cutoff gives an rms fluctuation strength $\delta \mu = A_0 \sqrt{\log( f_{\text{cutoff}} t_{m})}$, so
\begin{equation}
    \delta F=A_0 \sqrt{\log( f_{\text{cutoff}} t_{m})} /\alpha l.
\end{equation}

So, for a field fluctuation along $\hat{n}$, one has
\begin{equation}\label{eq:snr}
    \text{SNR} =\sqrt{\frac{2}{15}} \frac{\log (f_{\text{cutoff} } t_m)}{\sqrt{\text{arccot}(\kappa t_m)}}\frac{g^3}{\Delta^3} \left(\frac{8}{5}| \vec{\nabla}_F \omega_{vs} \cdot \hat{n}|\frac{A_0 }{l \alpha \kappa}\right)^2
\end{equation}
when substituting $n= \frac{2}{15}n_{\text{crit}}$ in for the noise, $N$, defined in Eq.~\eqref{eq:noisepower}.
%Additionally for the derivation of the sensitivity $\mathcal{S}$ to stay valid $\gamma$ and $\kappa$ have to be small compared to $\Delta$.

\begin{table}[]
    \centering
    \begin{tabular}{l l l}\hline
       Parameter  & Symbol& Value \\ \hline
       Charge noise strength &$A_0$ & 1.4 $\mu$eV/$\sqrt{\text{Hz}}$~\cite{connors_low-frequency_2019,struck_low-frequency_2020,freeman_comparison_2016,connors_charge-noise_2022,paquelet_wuetz_reducing_2023, esposti_low_2023}\\
       High frequency noise cutoff &$f_{\text{cutoff}}$ &100 MHz~\cite{connors_charge-noise_2022}\\
       Confinement energy &$E_o$ & 0.5 meV\\
       Lever arm & $\alpha$ & 0.08 eV/V~\cite{connors_low-frequency_2019}\\
       Dot distance from gate & $l$& 100 nm \\
       Valley decay rate& $\gamma$ & 2 MHz~\cite{boross_control_2016,penthorn_direct_2020}\\
       Resonator zero point voltage& $V_{rms}$ & 20 $\mu$V~\cite{samkharadze_high-kinetic-inductance_2016} \\
       Resonator frequency  & $\omega_c$ & 4 GHz~\cite{kandel_high-impedance_2023} \\
       Resonator decay rate&  $\kappa$& 0.5 MHz~\cite{samkharadze_high-kinetic-inductance_2016}\\
       Transmission line resistance & $R$ & 50 $\Omega$
       %& & \\
    \end{tabular}
    \caption{Parameter values used to calculate the measurement times for unity SNR.}
    \label{tab:params}
\end{table}

\section{Effect of the interface, alloy disorder, and uniform Ge in the well }\label{sec:tightbinding}
Several parameters in Eq.~\eqref{eq:snr} strongly depend on the details of the Si/SiGe heterostructure. The resonator-valley coupling strength $g$ is determined by the intervalley dipole moment $\vec{r}_{01}$, and the gradient of the valley splitting to electric field fluctuations, $\vec{\nabla}_F \omega_{vs} $, can be estimated as $e(\vec{r}_{11}-\vec{r}_{00})/\hbar$, the intravalley dipole moment splitting. Also, recall that $\Delta=\omega_{vs}-\omega_c$. Below, we calculate the valley splitting $\omega_{vs}$ and these dipole moments via tight-binding simulation, which along with the parameter values listed in Table \ref{tab:params} gives us all the information needed to evaluate Eq.~\eqref{eq:snr}. 

\subsection{Simulation of valley splitting, intervalley and intravalley dipole moments}\label{sec:TBsims}
To achieve fast \textit{in situ} charge noise measurements in this framework, it is desirable to have intervalley and intravalley dipole moments that are large, and valley splittings that are large compared to the thermal energy but otherwise as small as possible. Due to the strong quantum well confinement, it is difficult to achieve large dipole moments in the heterostructure growth direction $z$. On the other hand, large dipole moments can be achieved in lateral directions, especially in the presence of sharp interfaces and interface roughness~\cite{Hosseinkhani:2021p085309}, which has been considered to be a ubiquitous source of variability in these systems. 

\begin{figure*} 
    \centering 
    \includegraphics[width=1.0 \linewidth]{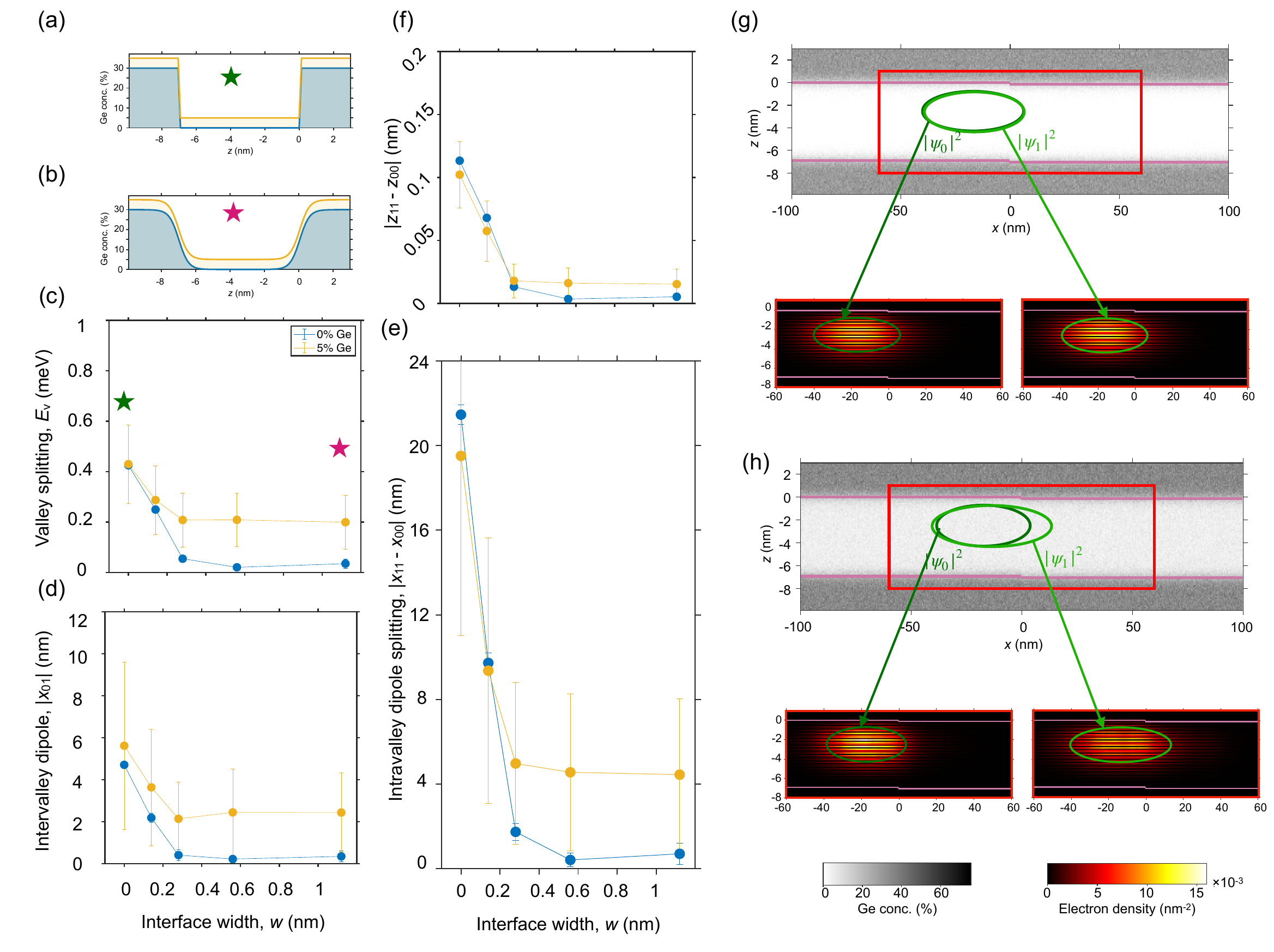}
    \caption{Simulation of valley parameters. (a-b)  Germanium concentration profiles of quantum wells with (yellow) and without (blue) additional Ge inside the well, for sharp ($w=0$) and wide ($w=1.12$ nm) interfaces, respectively. The interfaces have sigmoidal profiles. (c) Valley splitting rapidly decays with increasing interface width in conventional quantum wells (blue data). Adding a small amount of Ge inside the well increases average valley splitting and its variability (yellow data). (d) Intervalley dipole moment determines the strength of the resonator coupling and follows a similar trend. (e,f) The splitting between intravalley dipole moments determines the sensitivity of the valley splitting to electric field fluctuations. The $x$ component of this quantity (e) is about two orders of magnitude larger than the $z$ component (f). (g-h) Ge concentration of a conventional quantum well (g) and a quantum well with added Ge (h), with a wide interface ($w=1.12$ nm) and an interface step at $x=0$. Dark and light green ovals denote the localization of the dot electron in the ground and first-excited states, respectively and are obtained by fitting Gaussian envelopes to electron densities. Electron densities themselves, inside the region marked indicated by red boundaries, are also shown separately below.}  
    \label{fig:TBsims}
\end{figure*}
\begin{figure*} 
    \centering 
    \includegraphics[width=1.0 \linewidth]{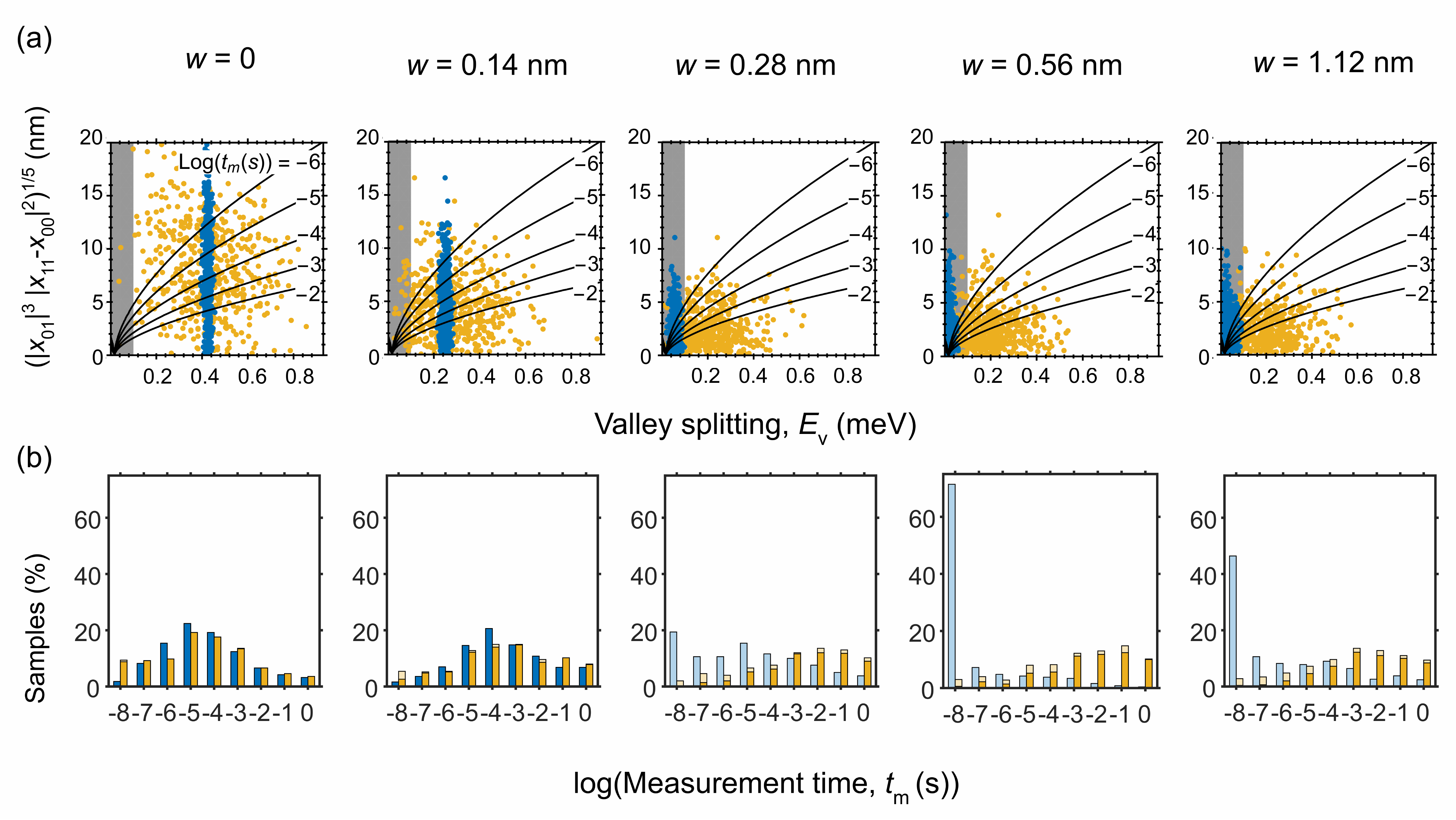}
    \caption{Distribution of valley splitting and dipole moment contribution, and the corresponding measurement times for the data shown in Fig.~\ref{fig:TBsims}(c-f). (a) Scatter plots show the valley splittings and the geometric means of the intervalley dipole moment and the intravalley dipole moment splitting obtained from 500 tight-binding simulations for five different interface width values, for a conventional quantum well (blue color) and a quantum well with 5\% Ge (yellow points). Corresponding measurement times are indicated by the contour lines in logarithmic scale. The shaded region indicates $E_v\leq 0.1 $ meV, valley splitting values that are considered too small for qubit operation. (b) Histograms of the data in (a) where the bar labeled by e.g. -3 indicates the samples with measurement times between $10^{-4}$s and $10^{-3}$s. Data that falls into the shaded region is indicated by lighter color bars. }  \label{fig:distr}
\end{figure*}

Recent studies have shown, however, that Si/SiGe interfaces are rather wide and random alloy disorder is the dominant source of variability~\cite{PaqueletWuetz:2022p7730, lima_interface_2023}. In other words, sharp features such as sharp interfaces or interface steps are less likely to play an important role in the physics of realistic devices. As a result, there have been several alternative proposals to enhance valley splitting~\cite{Losert:2023p125405}, which is otherwise known to favor sharp interfaces. One of the simplest of these proposals, which are commonly based on the idea of increasing the overlap of the electron wave function with the Ge atoms, consists of adding a constant and small concentration of Ge into the well. Ge concentration profiles corresponding to sharp and wide interfaces are illustrated in Fig.~\ref{fig:TBsims}(a) and (b), respectively, where blue markers indicate a conventional $\mathrm{Si/Si_{0.7}Ge_{0.3}}$ quantum well and yellow indicates a $\mathrm{Si_{0.95}Ge_{0.05}/Si_{0.65}Ge_{0.35}}$ quantum well. The implication for the valley charge noise probe is the following: while the desirable large dipole moments are difficult to achieve by interface steps, Ge doped quantum wells can induce dipoles due to the random distribution of Ge atoms in addition to their originally proposed goal of enhancing valley splitting. 

To quantify these implications, tight-binding simulations are performed on a set of 500 realizations of alloy disorder for each input parameter set in the presence of a single atomic interface step positioned at the center of the parabolic confinement potential with the isotropic orbital energy of 0.5 meV. The simulation approach is based on Ref.~\cite{Losert:2023p125405} and additional details are given in Appendix~\ref{sec:TBdetails}. Figures~\ref{fig:TBsims}(c-f) show the averages and standard deviations of key quantities such as valley splitting (c), intervalley dipole moment (d), and the $x$ (e) and $z$ (f) components of intravalley dipole moment splitting as a function of interface width $w$, for a conventional quantum well (blue color) and a quantum well with 5\% added Ge (yellow color). Similar trends are observed in all key quantities: (i) rapid decay with increasing interface width in conventional quantum wells, (ii) increase in average values and variability with an additional 5\% Ge, in comparison to the conventional case. It is important to note, however, that the $x$ component of the intravalley dipole moment splitting is about two orders of magnitude larger than its $z$ component, indicating that valley splitting is much more sensitive to lateral electric field fluctuations. This is expected because the lateral confinement is much weaker than the vertical confinement.

To further illustrate the dipole-inducing effect of added Ge inside the well in the presence of a wide interface, electron localizations corresponding to the two lowest eigenstates in typical alloy-disordered realizations of a conventional quantum well and a quantum well with added Ge are shown in Figs.~\ref{fig:TBsims}(g) and (h), respectively. Although there is an interface step at the center of the dot, both the ground and the first-excited state electrons essentially coincide in space in a conventional quantum well, as indicated by the boundaries (green colors) in Fig.~\ref{fig:TBsims}(g). Figure~\ref{fig:TBsims}(h) shows, on the other hand, that with the addition of 5\% Ge into the well it is possible to induce intervalley dipole moment and intravalley dipole moment splitting, as indicated by the boundaries (green colors) marking the ground and first-excited state electrons that are significantly offset.

\subsection{Measurement times}\label{sec:tm}
A contour plot of the measurement time producing unity SNR is shown in Fig.~\ref{fig:distr}(a) as a function of valley splitting and $\sqrt[5]{(x_{11}-x_{00})^2 x_{01}^3}$, which is proportional to the SNR term $(|\nabla_F \omega_{vs}|^2 g^3)$ in Eq.~\eqref{eq:snr}. The dots in Fig.~\ref{fig:distr}(a) represent the 500 tight-binding simulations at each interface width $w$ for both no Ge (blue) in the well and 5\% Ge in the well (yellow). Fig.~\ref{fig:distr}(b) shows a histogram of measurement times. For real-time recalibration or feedforward using mid-circuit measurements to be feasible, the measurement times should be faster than the qubit decoherence time $T_2$ which can be well over 1 ms~\cite{veldhorst_addressable_2014,yoneda_quantum-dot_2018}. For interfaces narrower than a couple of angstroms, we find that measurement times of below 1 ms can be obtained in a large percentage of samples for both the conventional well case and the 5\% Ge case, while for wider interfaces, even in the few samples where the measurement times are good, the valley splitting for the conventional well is below 100 $\mu$eV as indicated by the shaded region in Fig.~\ref{fig:distr}(a) and lighter histogram bars in Fig.~\ref{fig:distr}(b). Valley splitting below 100 $\mu$eV is undesirable since thermal occupation of the valley states becomes a concern with $k_b T=9 \, \mu$eV at 100 mK~\cite{mcjunkin_sige_2022,hollmann_large_2020,zhang_giant_2020,yang_spin-valley_2013}. Thus adding 5\% Ge in the well is desirable. For a realistic interface width such as $0.56$nm~\cite{esposti_low_2023}, adding 5\% Ge in the well results in 104 out of 500 samples having a measurement time below 1ms and a valley splitting above 100$\mu$eV. It may be further possible to tune other samples below 1ms by moving the dot position a little because the dipole moments and valley splitting could depend on the position of the dot. Alternative heterostructure designs could also be considered to achieve similar results.

\section{Conclusions} \label{sec:conclusions}
We have theoretically shown that in the presence of alloy disorder in the well of a Si/SiGe quantum dot the resulting intervalley dipole allows fast microwave reflectometry of lateral electric field fluctuations via changes in the valley splitting.
The coupled resonator-valley system then forms an \textit{in situ} probe of the charge noise while leaving the spin degree of freedom untouched, as long as one is not near an anti-crossing of spin and valley levels where there is a spin-valley relaxation ``hot spot''~\cite{yang_spin-valley_2013,hao_electron_2014,HuangPRB2014}. 

In principle, this probe could also be implemented in SiMOS systems, although there the confinement energy is typically larger due to the smaller vertical distance between the top gates and the dot~\cite{Saraiva_Materials_2022}. Since the intervalley dipole is induced by mixing in some excited orbital state character, it is clearly limited by the dot size~\cite{Gamble2013} (see Appendix \ref{sec:TBdetails} for further discussion of the scaling with confinement energy), so smaller dots would tend to have a weaker resonator-valley coupling. Ref.~\cite{yang_spin-valley_2013} inferred an intervalley dipole of only 1 nm. On the other hand, the interface is effectively sharper for SiMOS on the length scale of the dot, so an interface step may be more effective in producing an intervalley dipole than in Si/SiGe.

In conclusion, the valley degree of freedom in silicon-based qubits, long an annoyance, can be leveraged as an effective built-in sensor. This opens the possibility to monitor the dynamics of the charge noise in real time during the implementation of a spin qubit quantum circuit, allowing for closed-loop control, on-the-fly device recalibration, or feedforward to the subsequent structure of the circuit.

\section*{Acknowledgements} \label{sec:acknowledgements}
The authors acknowledge support from the Army Research Office (ARO) under Grant Number W911NF-23-1-0115, and thank John Nichol, Merritt Losert, Christopher Anderson, Susan Coppersmith, and Mark Friesen for useful discussions. 

\appendix
\section{Details of tight-binding simulations}\label{sec:TBdetails}
In this work, a two-band tight-binding model~\cite{Boykin:2004p115} is used as in Ref.~\cite{Losert:2023p125405}. While it is possible to account for the effect of alloy disorder on the valley splitting using a one-dimensional model along $z$ ([001] direction of the crystal), where the lateral distribution of Ge atoms is accounted for by performing a weighted averaging, a two-dimensional model is preferred here to observe the effect of an interface step and to be able to calculate the $x$ components of the dipole moments. In this minimal model, nearest and next-nearest neighbor hopping parameters along $z$, given by $t_1=0.68$ eV and $t_2=0.61$ eV, are chosen to reproduce the lateral effective mass $m_l=0.916m_e$ and the reciprocal space location of the conduction band minima $k=\pm0.82(2\pi/a_0)$ for bulk Si, where the lattice spacing is set to $\Delta_z=a_0/4$. 
%Note that only two of the six conduction valleys are relevant for the low-temperature physics as the remaining four are lifted in energy due to strain. 
The nearest-neighbor hopping term in $x$, $t_3=-1.36$ eV, is set to yield the transverse effective mass $m_t=0.19m_e$ for a grid spacing of $\Delta_x=a_0/\sqrt{2}$. 

\begin{figure} 
    \centering 
    \includegraphics[width=1.0 \linewidth]{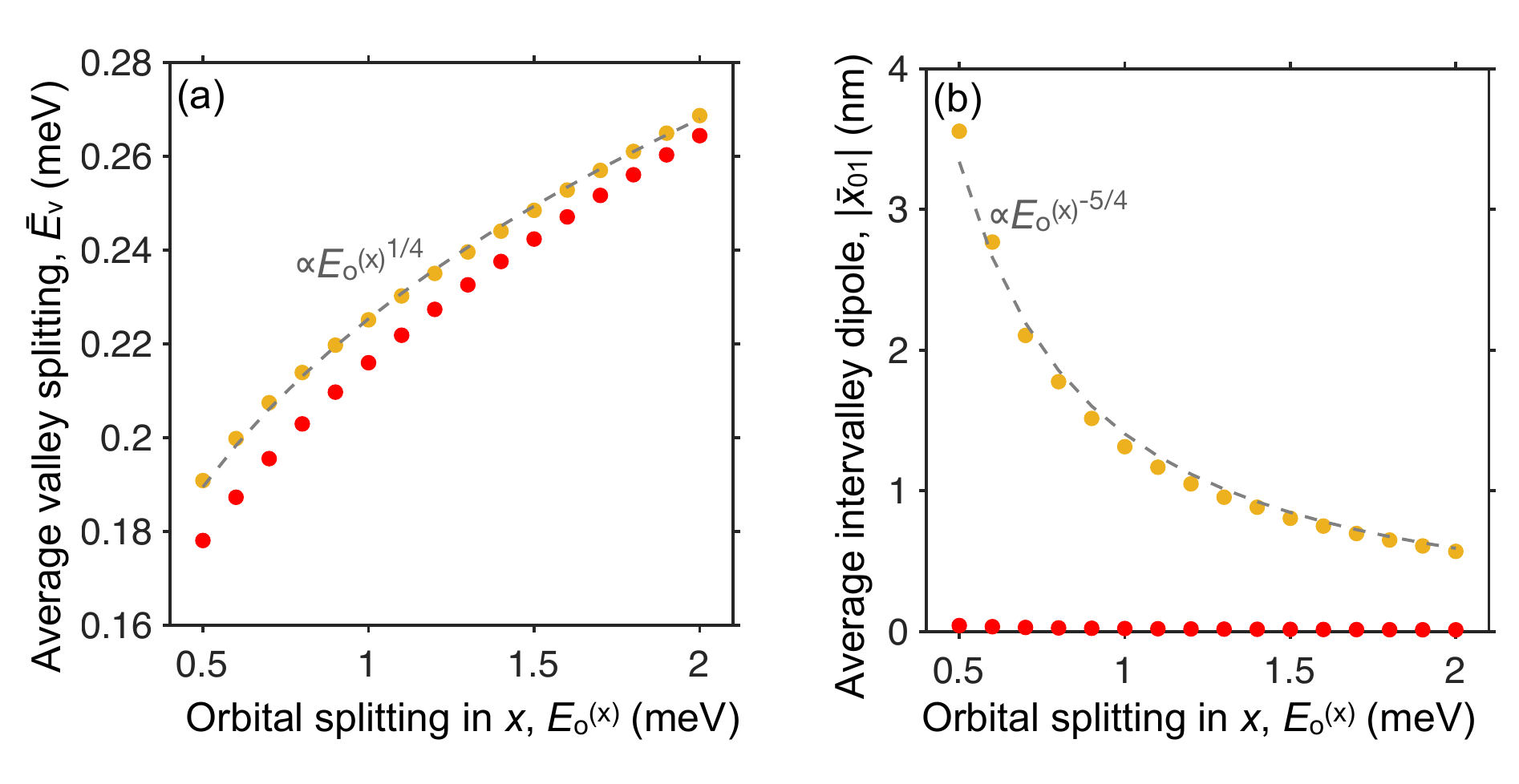}
    \caption{Effect of dot size and lateral distribution of Ge atoms on (a) average valley splitting and (b) average intervalley dipole moment, where averaging is performed over 500 random alloy realizations While yellow points are based on samples that have purely random lateral distribution of Ge atoms, red points correspond to the samples with the random Ge concentration profiles artificially made symmetric around $x=0$. Gray-dashed lines are the fitted curves (to the yellow points) with the indicated forms that are discussed in the text.}  
    \label{fig:dotSize}
\end{figure}

Effects of the randomly placed Ge atoms and the gate-induced electric fields are captured by the on-site energies in the tight-binding model. Interfaces are modeled with sigmoidal Ge profiles. Simulations are performed in a 200 nm $\times$ 13 nm domain in the $xz$ plane and the quantum well size is set to 7 nm. The vertical electric field is set to 5 MV/m and the lateral confinement potential is chosen to be isotropic and parabolic, for simplicity, with the orbital energy $E_o$ of 0.5 meV. This orbital energy implies a quantum dot confinement that is on the weaker side of the spectrum and is chosen to yield desirable parameters (i.e. smaller valley splittings and larger dipoles) for the valley probe in the presence of 5\% Ge in the well. Particularly, larger dots lead to smaller average valley splittings ($\bar{E}_v$) and larger average dipole moments ($\bar{x}_{ij}$). The scaling of $\bar{E}_v=2\overline{|\bra{\psi_+}U_{\mathrm{qw}}\ket{\psi_-}|}$, where $\psi_{\pm}$ represent the degenerate valley states and $U_{\mathrm{qw}}$ is the quantum well potential, with the orbital energy is found to be $\bar{E}_v \sim (E_o^{(x)} E_o^{(y)})^{1/4}$ with the help of effective mass theory and statistical tools~\cite{Losert:2023p125405}. In a similar spirit, the scaling of the dipole moments can be understood. For instance, to induce intervalley dipole moment $x_{01}$, the valley states in the lowest orbital need to mix with the valley states in the first-excited $x$ orbital. This mixing, on average, is proportional to $\overline{\bra{\psi_{g,+}}U_{\mathrm{qw}}\ket{\psi_{e_x,-}}}/E_o^{(x)} \sim (E_o^{(x)})^{-3/4}$, where subscripts $g$ and $e_x$ denote the ground and first-excited $x$ orbitals of a 2D harmonic oscillator. Noting that $\bra{g} x \ket{e_x} = \hbar/\sqrt{m^*E_o^{(x)}}$, the scaling relation $\bar{x}_{01}\sim(E_o^{(x)})^{-5/4}$ is found. Figure~\ref{fig:dotSize} shows (a) $\bar{E}_v$ and (b) $\bar{x}_{01}$, in yellow color, as functions of $E_o^{(x)}$ when $E_o^{(y)}$ is kept constant at 2 meV, along with the fits to the leading terms in $E_o^{(x)}$ indicated by gray-dashed lines.

% \begin{figure}
%     \centering 
%     \includegraphics[width=1.0 \linewidth]{fig3.pdf}
%     \caption{Mechanism of alloy disorder induced intervalley dipole. Simulations are performed with random germanium distributions that are made symmetric around $x$=0. (a) For the valley splitting layer to layer change in the Ge distribution is important. This is why the problem can be reduced to 1D in the absence of interface steps and the valley splitting is insensitive to specific distribution of Ge atoms in the $xy$ plane  (b) Symmetry breaking in the $x$ direction is necessary to get nonzero intervalley dipole. Nonzero averages for the first two points indicate that the ground state gap is likely dominated by orbital excitations rather than valley excitations in some of these weakly confined dots.}  
%      \label{fig:dotSym}
% \end{figure}
\section{Numerical master equation simulations}\label{sec:NumericalMaster}
We numerically solved the master equation Eq.~\eqref{eq:mastereqn1}, with the addition of the orbitals as described in Sec.~\ref{sec:orbitalseffects}, to confirm the validity of the analytical expression for the sensitivity, $\mathcal{S}$, in Eq.~\eqref{eq:sense} derived using input-output theory in the dispersive limit $g/\Delta\ll 1$. In addition to the parameters from Table~\ref{tab:params}, we chose $\omega_{vs}=300~\mu$eV and $|x_{01}|=5$~nm as they are representative of the tight-binding simulations in Sec.~\ref{sec:tightbinding}. We truncated the Fock space of the resonator to a maximum occupation of six photons. Additionally, we used the orbital energy of $\omega_o =0.5$~meV from in Sec.~\ref{sec:tightbinding}. The interorbital dipole moment was estimated as the dipole moment between the ground and first excited Fock-Darwin states which is $x_{02}\approx\frac{1}{2}\sqrt{\frac{\hbar}{m^* \omega_{o}}} \approx 15$~nm where $m^*=0.19 m_e$ is the electrons effective mass in silicon and $m_e$ is the electron mass. Both the orbital and valley system are in the dispersive regime since $g/\Delta\approx 0.0018$ and $g_{o}/\Delta_{o}\approx 0.0031$. 

\begin{figure} 
    \centering 
    \includegraphics[width=1.0 \linewidth]{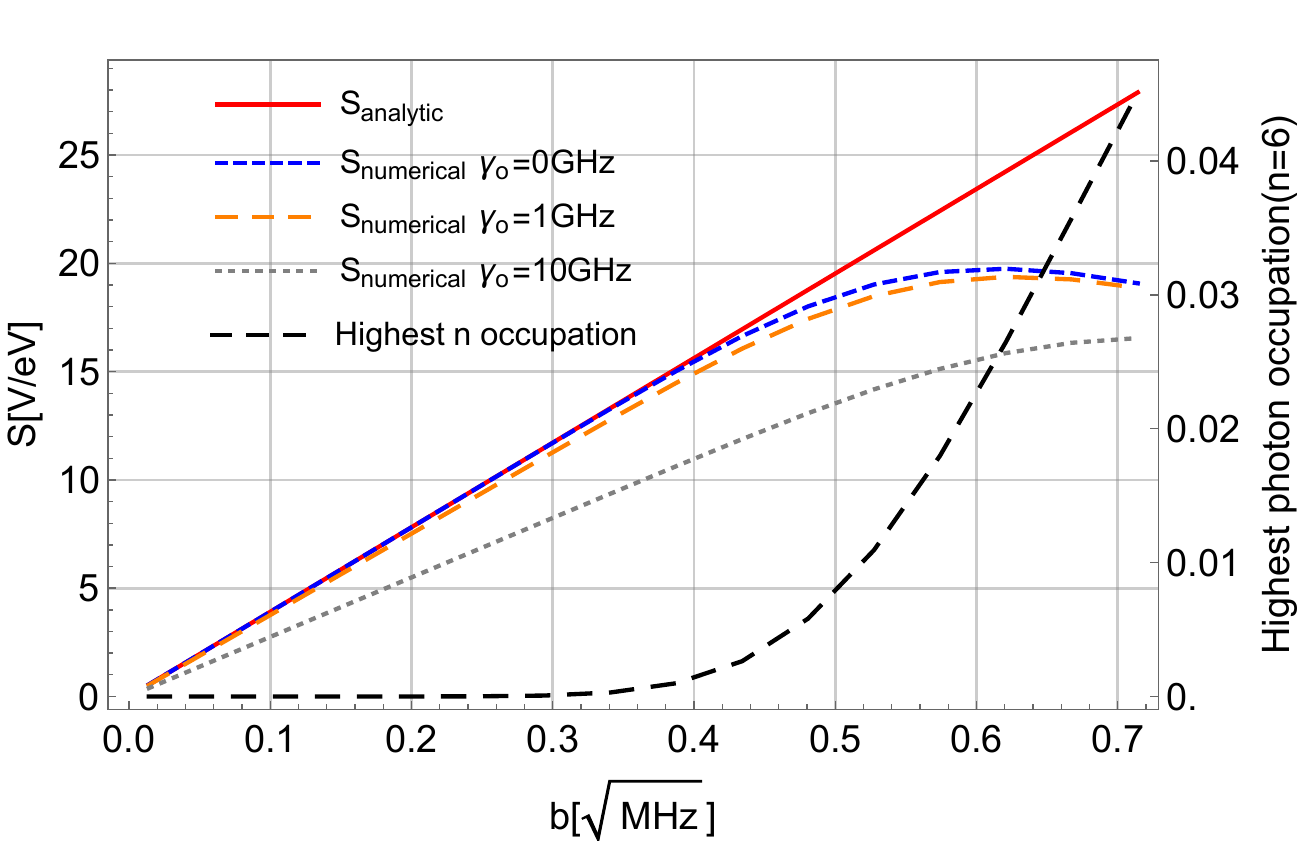}
    \caption{Sensitivity of the out-of-phase voltage, $\mathcal{S}$, calculated numerically(blue) and using Eq.~\eqref{eq:sense} for different probe field strengths, $b$. Additionally, the occupation of the highest photon level $n=6$ is shown as a function of $b$ as well which indicates that our numerical simulation becomes invalid after $b=0.3\sqrt{\text{MHz}}$ .}  
    \label{fig:mastereqn}
\end{figure} 
\begin{figure} 
    \centering 
    \includegraphics[width=1.0 \linewidth]{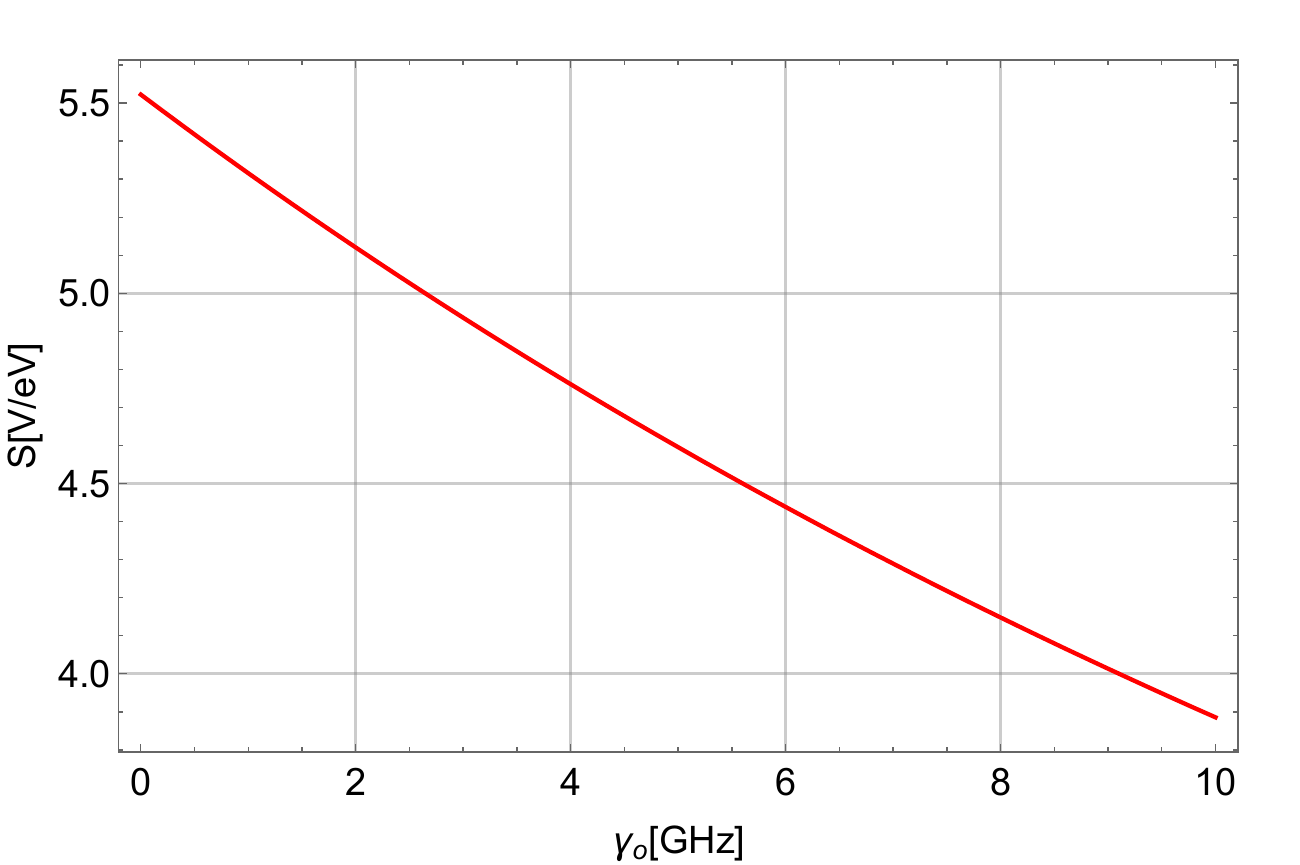}
    \caption{Sensitivity of the out-of-phase voltage, $\mathcal{S}$, calculated numerically as a function of orbital decay rate, $\gamma_o$, at probe field strength $b=0.015\sqrt{\text{MHz}}$.}  
    \label{fig:mastereqngamma}
\end{figure} 

Figure~\ref{fig:mastereqn} shows the numerically computed sensitivity, $\mathcal{S}_{\text{numerical}}$, as a function of probe strength for different orbital decay rates. The $\gamma_o=0$ numerical result matches the analytical value from Eq.~\eqref{eq:sense}, $\mathcal{S}_{\text{analytic}}$, shown in red in Fig.~\ref{fig:mastereqn}, for probe field strengths $b<0.4\sqrt{\text{MHz}}$. For stronger probing the numerical calculation in our truncated space is unreliable since the occupation of the highest photon level, shown in black in Fig.~\ref{fig:mastereqn}, becomes non-negligible and further excitations above $n=6$ map unphysically to the null vector. As shown in Eq.~\eqref{eq:napprox}, the number of photons in the resonator while probing on resonance grows quadratically with $b$, so we are limited in the range of probe strength we can numerically calculate accurately, whereas input-output theory has no such limitation as long as $n\ll n_{\text{crit}}$. However, the numerical approach allows us to confirm the validity of the various approximations made in our analytical result in the low-power regime, as well as to consider the additional effect of orbital relaxation.

Fig.~\ref{fig:mastereqngamma} shows the numerically calculated sensitivity at $b=0.015\sqrt{\text{MHz}}$ for different values of $\gamma_o$. At $\gamma_o=1$~GHz and $\gamma_o=10$~GHz the sensitivity decreases by $\approx$ 4\% and $\approx$ 30\% respectively from the value at $\gamma_o=0$.

The heating caused by the decay of the orbital state into phonons, $P_{o}$, was estimated using 
\begin{equation}
    P_o=\frac{1+\langle \tau_z \rangle_{ss}}{2}  \gamma_o  \omega_o \approx \frac{2 \Delta^2 g_o^2 \gamma_o \omega_o}{15 g^2 \Delta_o^2}
\end{equation}
where the occupation of the excited orbital $(1+\langle \tau_z \rangle_{ss})/2\approx0.09$ was estimated using Eq.~\eqref{eq:steadysigmaz} at probe power $b=\sqrt{\kappa n_{\text{crit}}/30}$. For an orbital decay rate $\gamma_o=1$~GHz we get $P_o=7.2 \times 10^{-14}$~W.

\section{Higher order expansion of input-output theory}\label{sec:appenFourthorder}
\begin{figure} 
    \centering 
    \includegraphics[width=1.0 \linewidth]{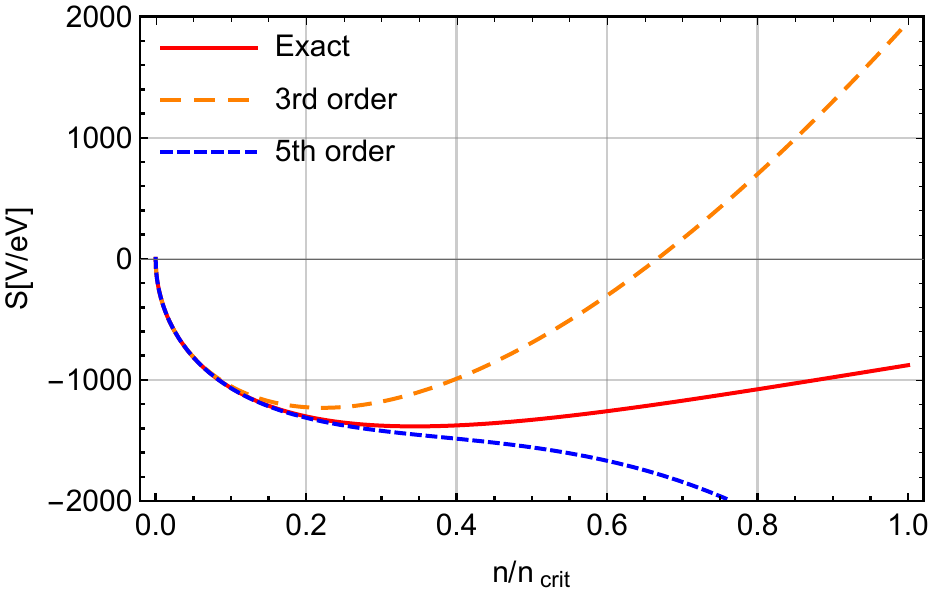}
    \caption{Sensitivities calculated exactly from input-output theory (red solid line), using Eq.~\eqref{eq:sense} (orange dashed line) and Eq.~\eqref{eq:sense5ord} (blue small dashed line) as a function of the number of photons in the cavity.}  
    \label{fig:higherBexpansion}
\end{figure}

In this section, we verify that truncating the expansion of $\langle \sigma_{z} \rangle_{ss}$ at second order in $b/\sqrt{\kappa}$ is accurate enough for our sensitivity found in Eq.~\eqref{eq:sense} and we verify that taking $\gamma_{eff}\rightarrow 0$ does not affect our order of magnitude measurement time calculations. We first calculate the sensitivity by solving Eqs.~\eqref{eq:dadt}-\eqref{eq:dsigma-dt} exactly while including $\gamma_{eff}$ and the resonance frequency from Eq.~\eqref{eq:higherOmegaRes}, which results in an analytic expression which is too complicated to print here. We then compare the plot of the exact result to simpler approximate expressions derived up to different orders of $b/\sqrt{\kappa}$ with $\gamma_{eff}\rightarrow 0$. Using the representative values taken in Appendix \ref{sec:NumericalMaster}, the exact result is plotted in Fig.~\ref{fig:higherBexpansion} as a red solid line. The sensitivity as calculated using Eq.~\eqref{eq:sense} is also shown in Fig.~\ref{fig:higherBexpansion} as an orange dashed line. From Fig.~\ref{fig:higherBexpansion} it is clear that Eq.~\eqref{eq:sense} is only valid to about 0.1$n_{\text{crit}}$ and underestimates the magnitude of the sensitivity for $n< 0.85 n_{\text{crit}}$. 

To check how much the approximation improves by going to a higher order expansion, we further expand $\langle \sigma_{z} \rangle_{ss}$ up to fourth order in $b/\sqrt{\kappa}$ (which is proportional to $n$ through Eq.~\eqref{eq:napprox}). 
The added fourth-order term is
\begin{equation}
   \frac{64 g^4 \kappa^2(4g^4 (\omega_{vs}-\omega_{res})^2(4 (\omega_c-\omega_{res})^2+\kappa^2))}{(4(g^2-(\omega_c-\omega_{res})(\omega_{vs}-\omega_{res}))^2+(\omega_{vs}-\omega_{res})^2\kappa^2)^3} b^4. 
\end{equation}
Combining this with the lower order terms in Eq.~\eqref{eq:steadysigmaz} we find a new approximation
for $\langle \sigma_{z} \rangle_{ss}$, 
\begin{equation}
    \begin{split}
        \langle\sigma_{z}\rangle_{ss}\approx &-1+\frac{8b^2+\Delta^4\kappa^2}{256 b^4 g^8\Delta^6\kappa^4}\\
        &-\frac{64 b^4 g^4\Delta^8 \kappa^6(256 b^4 g^8-64b^2 g^6\Delta^2\kappa+\Delta^6 \kappa^4)}{(256 b^4 g^8+\Delta^6\kappa^4)^3},
    \end{split}
\end{equation}
which was found by taking $\omega_{vs}-\omega_{res} \rightarrow \Delta$ because the resonance frequency is close to $\omega_c$ and taking $\omega_c-\omega_{res} \rightarrow \frac{g^2}{\Delta}-\frac{8 g^4 b^2}{\Delta^3 \kappa}$ because the resonance frequency will be close to the lower order resonance frequency in Eq.~\eqref{eq:omegares} which at large $b$ is approximately $\omega_c-\frac{g^2}{\Delta}+\frac{8 g^4 b^2}{\Delta^3 \kappa}$. With this new value of $\langle\sigma_{z}\rangle_{ss}$ we also find a new resonance frequency using Eq.~\eqref{eq:omegadispexpan}
\begin{equation}\label{eq:higherOmegaRes}
    \begin{split}
        \omega_{res} &\approx \omega_c+\frac{g^2 \langle\sigma_{z}\rangle_{ss}}{\Delta}
        \\
        &\approx \omega_c-\frac{g^2}{\Delta}+\frac{8 b^2 g^4 \Delta^3\kappa^3}{256b^4g^8+\Delta^6\kappa^4}
        \\
        +& \frac{64 b^4 g^6\Delta^7\kappa^6(4g^4\Delta^4\kappa^2-\Delta^6\kappa^4-4(g^2\Delta^2\kappa-8b^2g^4)^2)}{(256b^4g^8+\Delta^6\kappa^4)^3}
    \end{split}
\end{equation}
Calculating the sensitivity, $\mathcal{S}$, by finding the steady state solutions of Eq.~\eqref{eq:dadt} -\eqref{eq:dsigma-dt} with the new values for $\langle\sigma_{z}\rangle_{ss}$ and $\omega_{res}$ yields a very complicated expression. To simplify this expression we Taylor expand it in $b/\sqrt{\kappa}$ around 0. If this is done up to the third order the resulting sensitivity, 
\begin{equation}\label{eq:sense3ord}
   \mathcal{S} = \frac{4b \sqrt{2 \hbar \omega_c R }}{\kappa}\left(-\frac{g^2}{\Delta^2}+\frac{24g^4b^2}{\kappa \Delta^4}  + \mathcal{O}\left(\frac{b^4}{\kappa^2}\right) \right),
\end{equation}
is the same as Eq.~\eqref{eq:sense} after dropping the $bg^4/\Delta^4$ term.
The fifth-order Taylor expansion of the new equation is
\begin{equation}\label{eq:sense5ord}
   \mathcal{S} = \frac{4b \sqrt{2 \hbar \omega_c R }}{\kappa}\left(-\frac{g^2}{\Delta^2}+\frac{24g^4b^2}{\kappa \Delta^4}-\frac{320g^6b^4}{\kappa^2 \Delta^6}  +  \mathcal{O}\left(\frac{b^6}{\kappa^{3}}\right) \right)
\end{equation}
which is shown in Fig.~\ref{fig:higherBexpansion} as a dashed blue line. This increases the range over which the equation is valid to about 0.3$n_{\text{crit}}$. 

However, from our SNR analysis in Section \ref{sec:SNR} we found the optimal SNR for Eq.~\eqref{eq:sense} to be around $2/15 \times n_{\text{crit}}\approx 0.13 n_{\text{crit}}$ which is where Eq.~\eqref{eq:sense} is still close to the exact sensitivity in Fig.~\ref{fig:higherBexpansion}. The point of optimal SNR from the exact results is at $\approx 0.16 n_{\text{crit}}$ where SNR$\propto \mathcal{S}^2/\sqrt{n}$ is at a maximum. This means Eq.~\eqref{eq:sense} is a slight underestimate for the maximum SNR but still sufficient for the dynamics we are interested in. 

Additionally, the approximate equations derived using $\gamma=0$ matched the exact sensitivity taking into account $\gamma_{eff}=\sqrt{2n}\kappa+\gamma$. Since they match closely this confirms that the backaction of the cavity on the system does not affect the order of magnitude of the measurement time calculations.

\bibliographystyle{apsrev4-1} 
\bibliography{refs} 
\end{document}